\documentclass{elsart}

\usepackage{natbib}

\usepackage{amssymb}

\begin{document}

\begin{frontmatter}


\title{Hypernovae and light dark matter as possible Galactic positron sources}
\author{S. Schanne, M. Cass\'e, J. Paul, B. Cordier}
\address{CEA Saclay, DSM/DAPNIA/Service d'Astrophysique, 91191 Gif sur Yvette, France.}

\title{}


\author{}

\address{}


\begin{abstract}

The electron-positron annihilation source in the Galactic center region has recently been observed with INTEGRAL/SPI, which shows that this 511 keV source is strong and its extension is consistent with the Galactic bulge geometry. 
The positron production rate, estimated to more than 10$^{43}$ per second, is very high and raises a challenging question about the nature of the Galactic positron source. 
Commonly considered astrophysical positron injectors, namely type Ia supernovae 
are rare events and fall short to explain the observed positron production rate. 
In this paper, we study the possibility of Galactic positron production by hypernovae events, exemplified by the recently observed SN2003dh/GRB030329, an asymmetric explosion of a Wolf-Rayet star associated with a gamma-ray burst. In these kinds of events, the ejected material becomes quickly transparent to positrons, which spread out in the interstellar medium.
Non radioactive processes, such as decays of heavy dark matter particles (neutralinos) predicted by most extensions of the standard model of particle physics, could also produce positrons as byproducts. However they are expected to be accompanied by a large flux of high-energy gamma-rays, which were not observed by EGRET and ground based Tcherenkov experiments. In this context we explore the possibility of direct positron production by annihilation of light dark matter particles.

\end{abstract}



\end{frontmatter}

\section{INTEGRAL/SPI observation of $e^+$ annihilation towards the Galactic center}
\label{}

	SPI, the spectrometer onboard INTEGRAL \citep{schanne2002,attie2003,roques2003,vedrenne2003}, ESA's gamma-ray satellite launched in October 2002, has recently reported its first results on the observation towards the Galactic center region of the 511 keV gamma-ray line emission resulting from e$^{+}$ e$^{-}$ annihilation \citep{jean2003}. 
	The flux $\Phi_{511}=$(0.99$^{+0.47}_{-0.21}$) $\times$ 10$^{-3}$ ph cm$^{-2}$ s$^{-1}$, is concentrated in a narrow gamma-ray line at an energy of 511.06$^{+0.17}_{-0.10}$ keV with an intrinsic line width of 2.95$^{+0.45}_{-0.51}$ keV (FWHM).
	The spatial shape of the 511 keV emission region \citep{jean2003,weidens2004} fits best a spherically symmetric 2D Gaussian distribution projected on the sky, centered on the Galactic center, with a width of 10$^\circ$ (FWHM) (ranging from 6$^\circ$ to 18$^\circ$ with 95\% C.L.).
	No significant emission from the Galactic disk has been detected by SPI.
	A comparison with earlier observations of the 511 keV source, in particular the CGRO/OSSE observation \citep{purcell97}, is presented in \cite{weidens2004}.

	If we assume that the 511 keV emission takes place near the Galactic center located at a distance $R_{o}=$8.0 kpc, then the spatial size of the emission region, $D=$1.4 kpc in diameter, is comparable in size with the Galactic bulge.
	From the observed flux, we infer a 511 keV-photon production rate of $L_{511}=$7.7$\times$10$^{42}$ ph s$^{-1}$.
	A 511 keV photon can either be produced by direct $e^{+}$ $e^{-}$ annihilation or by $e^{+}$ $e^{-}$ annihilation via a para-prositronium intermediate state.
	As shown by \cite{kinzer2001} based on the CGRO/OSSE observations, in a fraction $f_{Ps}=$0.93 of the cases the $e^{+}$ $e^{-}$ annihilation takes place after positronium formation in a warm ($T\sim$10$^4$ K) medium.
	Combining the SPI and OSSE measurements, we get a $e^{+}$ annihilation rate in the Galactic bulge as large as $L_{e^+}=$1.3$\times$10$^{43}$ $e^+$s$^{-1}$.
	If we assume a steady-state positron production/annihilation, the same number of positrons must be injected each second into the Galactic bulge.
 
	From these considerations can be raised the question about the nature of the source capable of injecting such a number of positrons near the Galactic center into the Galactic bulge, as well as the question of the medium onto which those positrons annihilate.


\section{Astrophysical candidates for the source of Galactic positrons}

	Among the candidate sources capable of producing 511~keV gamma-ray line emission by positron production, we review astrophysical source candidates.
	Positrons could be released into the Galactic bulge by compact objects such as accreting back holes/microquasars \citep{mirabel1992}, however those sources are not numerous enough and they are active only during short periods of time, such that they do not appear to be the main Galactic positron injectors (though a more detailed study of their positron production remains to be done).
	Low Mass X-ray binaries (LMXB) have also been proposed as candidates for positron production \citep{pranzos2004}.
	However their positron yield is still uncertain, and furthermore their distribution concentrates in the Galactic bulge and the Galactic disk as well, as shown by the RXTE all-sky survey in the 3$-$20 keV energy band \citep{revnivtsev2004} and the soft gamma-ray source survey in the central radian of the Galaxy by INTEGRAL/ISGRI \citep{lebrun2004}.
	If positrons produced by LMXB annihilate locally, the morphology of the 511 keV map detected by INTEGRAL/SPI should show a disk component, which does not seem to be the case.

	The most promising astrophysical Galactic positron sources are  radioactive ejecta (produced by nucleosynthesis sites) which can
release large amounts of positrons into the interstellar medium 
through their $\beta^+$ decay.
	The list of relevant isotopes comprises $^{22}$Na which is expected to be produced by novae; however their positron production rate is too small and the $^{22}$Na line is absent in the spectrum of the Galactic central region \citep{leising1998}.
	The $^{26}$Al isotope has been observed, however the flux of the associated 1809 keV line in the Galactic central radian, i.e. 3$\times$10$^{-4}$ ph cm$^{-2}$ s$^{-1}$ \citep{knodl1999}, and its narrow latitude profile are incompatible with the 511 keV flux and morphology. 
	For $^{44}$Ti the expected flux is even smaller (based on the measured $^{44}$Ti/$^{26}$Al flux ratios for the $^{44}$Ti sources known in the Galaxy: Vela and Cas A).
	The most prominent source of $e^+$ in the Galaxy could be the $^{56}$Co isotope which decays to $^{56}$Fe releasing a positron 19\% of the time. 
	$^{56}$Co is the decay product of $^{56}$Ni, copiously synthesized in supernova explosions.
	Each SN~II synthesizes about 0.1 M$\odot$ of $^{56}$Ni but the thick H envelope traps the positrons inside the ejecta.
	A SN~Ia produces about 0.6 M$\odot$ of $^{56}$Ni and from its thinner envelope about 3.3\% of the positrons (i.e. $\simeq$8$\times$10$^{52}$ $e^+$ per SN~Ia) are expected to escape \citep{milne1999}.
	In order to explain the observed positron annihilation rate by SNe~Ia alone (in a steady-state positron production/annihilation scenario) a mean SN~Ia rate of 0.6 per century is required in the Galactic bulge.
	This requirement is in excess by a factor of 10 compared to the estimated SN~Ia rate in the Galactic bulge, as shown in \cite{schanne2004}, using results from SN~Ia statistical studies in elliptical galaxies by \cite{cappellaro2003} applied to the case of the Galactic bulge, which is an elliptical galaxy embedded in our Galaxy (the bulge is formed by $\sim2\times10^{10}$M$\odot$ of old stars, with a coeval formation $\sim$10 Gyr ago and a velocity dispersion comparable to that of a small elliptical galaxy).
	We conclude therefore that a small fraction of SN~Ia events contribute to the injection of positrons into the Galactic bulge.


\section{Hypernovae as a source for Galactic positrons}
	
	Following the observation of the unusual type Ic supernova SN2003dh \citep{hjorth2003,stanek2003} associated to the gamma-ray burst GRB030329, we proposed recently \citep{casse2004a,schanne2004} that hypernovae, i.e. explosions of Wolf-Rayet stars producing strongly asymmetric ejections, could be an important source of Galactic positrons.
	Indeed, the light curve of SN2003dh has been modeled by \cite{woosley2003} as an asymmetric explosion of a 10 M$\odot$ rotating Wolf-Rayet star which synthesizes $M_{Ni}$=0.5 M$\odot$ of $^{56}$Ni.
	Within a cone whose half opening angle is 45$^\circ$, a mass $M$=1.2 M$\odot$ is ejected with a high kinetic energy $E=$1.25$\times$10$^{52}$ erg.
	Since the duration after which positrons leak-out from the ejecta scales like $t_{e^+} \propto M E^{-1/2}$, the escape of positrons arises much earlier for SN2003dh than for a typical SN~Ia for which a mass $M$=1.34~M$\odot$ (i.e. the total mass of the white dwarf in the progenitor binary system) is ejected with a much lower kinetic energy of $E$=1.17$\times$10$^{51}$~erg (model DD23C, \cite{milne2001}).
	The fact that the light curve of SN2003dh decreases faster than the one of a typical SN~Ia supports the analysis of an early positron leak-out for SN2003dh, in which case the ejecta are less efficiently heated up by the decay products of the radioactive $^{56}$Ni isotope, synthesized in similar amounts by both explosions.
	In the case of strong mixing of the $^{56}$Ni with the ejecta, virtually all the 2$\times$10$^{54}$ e$^+$ formed by the $^{56}$Co decay escape.
	In this case SN2003dh would release 25 times more e$^+$ than a SN~Ia.
	The positron annihilation rate observed by SPI would be explained by SN2003dh-like hypernovae alone, if their average occurrence was of the order of 0.02 per century.

	The Galactic bulge is formed of very old stars and a very low density gas, where core collapse supernovae (and hence hypernovae) are not expected to take place.
	On the contrary, they are very likely to occur in the Galactic nuclear zone which is a very active and massive star forming region, in shape of a disk along the Galactic plane (450 pc in diameter, 50 pc in thickness), located around the Galactic center, composed of young massive main sequence stars, huge molecular clouds with a very clumpy distribution and a low volume filling factor, surrounded by a bath of cold atomic gas
\citep{launhardt2002,gusten2004}.
	Infra-red maps of the Galactic bulge by MSX in the 8~$\mu$m band \citep{bland2003} show the presence of dust in a region comparable in size with the Galactic nuclear zone.
	This dust could be accompanied by the presence of a warm and low density gas, in which positrons would form a significant fraction of positronium before annihilation, and produce a narrow 511 keV line.
	We propose therefore to compare this infra-red map with the annihilation map observed by SPI.
	Positrons produced in the Galactic nuclear zone by hypernova explosions could also leak out into the very low-density gas of Galactic bulge, where their stopping time would be much longer ($>$10 Myr) and where they would accumulate before annihilation.
	Matter propagation out of the nuclear zone into the Galactic bulge could take place by the means of a bipolar wind detected at large scales \citep{bland2003}.
	In this case, the positrons would fill-up the Galactic bulge and the morphology of the positron annihilation map would not allow anymore to trace the location of the positron source unambiguously, but rather give the distribution of the annihilation medium.
	Hypernovae are expected to occur in the Galactic disk as well, but due to the disk thinness ($\sim$100 pc in height), $e^+$ produced by disk-hypernovae are likely to leak out of the disk without in-situ annihilation.

	A very first and crude hypernova rate estimate in the Galactic nuclear zone \citep{schanne2004}, which has to be refined as more observations become available, shows that the overall SN~Ic rate in the Galactic nuclear zone is about 0.02 per century, while the rate of SN2003dh-like hypernova events (able to release large amounts of positrons) is only about 1\% of this rate, which would be unfavorable for hypernovae being a dominant positron injection candidate under the steady-state hypothesis.
	In a non steady scenario the starburst activity, which took place about 7 Myr ago in the Galactic center region and which may be recurrent on longer time scales \citep{bland2003}, could have produced positrons through SN2003dh-like hypernova events.
	Those positrons could have been injected into the Galactic bulge, where they are still present and annihilate today at a rate related to the properties  of the annihilation medium they encounter.


\section{Light dark matter annihilation as a possible positron source}

	In the absence of a solid astrophysical candidate for the observed positron source, it has been considered that the galactic 511 keV line emission is of non nuclear origin, and could open a door to new physics and create a link with one of the most intriguing questions of modern cosmology, namely the quest for the origin of the dark matter observed in galactic halos.

	One of the most popular dark matter candidates is the neutralino, the lightest supersymmetric particle, introduced in extensions of the standard model of particle physics, predicted with a high mass value ranging from 50 to 500 GeV/c$^2$.
	As a stable weakly interacting massive particle (WIMP), it would concentrate in deep gravitational potential wells, among which is the Galactic center region.
	The neutralino could be its own antiparticle, and therefore annihilate  where it concentrates.
	In this annihilation process, positrons would be produced, which in turn would annihilate and produce a 511 keV distribution centered on the Galactic center, as observed.

	The neutralino annihilation process is expected to occur via $W$ and $Z$ boson exchange, and produces quark-antiquark pairs, decaying into pion final states.
	Neutral and charged pions are expected to be produced in similar amounts, resulting in a similar number of high-energy photons and positrons produced.
	However, using the CGRO/EGRET data \citep{hunter1997}, the high-energy photon flux in the Galactic bulge has been constrained to $\Phi(\gamma>$100 MeV$)<8\times10^{-7}$ ph s$^{-1}$ cm$^{-2}$ by \cite{casse2004c}, which is too low compared to the measured positron annihilation rate.
	Furthermore, the high-energy positrons produced in the neutralino annihilation in the Galactic center region must slow down to thermal energies before they can form the observed positronium intermediate state with subsequent decay and production of the narrow 511 keV line.
	The processes involved (Bremsstrahlung on atoms of the insterstellar medium, inverse Compton scattering on optical photons produced by the starlight, and synchrotron emission in magnetic fields) would produce a large number of high-energy photons, contradicting once more the EGRET observation and the fact that the energy radiated in such a way by positrons injected at high energy (e.g. 1 GeV) would by far exceed the bolometric luminosity of the Galactic bulge.
	In summary, high energy particles such as annihilating neutralinos are not the observed positron source.

	Therefore a new scenario involving a new kind of light dark matter particles (LDMP) has been proposed \citep{fayet2004,boehm2004,casse2004b}.
	The LDMP, being their own anti-particles, annihilate into $e^+ e^-$ pairs, which in turn produce the observed 511 keV signature.
	The dark-matter mass density being fixed, a low LDMP mass (ranging from 10 to 100 MeV) implies a high LDMP number density and therefore a high annihilation rate, producing the high positron production rate observed.
	The positrons are produced with low energy which avoids any problematic high-energy photon production during slowdown.
	The small flux of observed high-energy photons (which can also be produced by internal bremsstrahlung during LDMP annihilation) furthermore constrains the LDMP mass \citep{beacom2004}.
	Without a direct coupling to the $Z$ boson, the LDMP would have escaped detection in accelerator based particle physics experiments.
	The LDMP annihilation cross section can be written with a dependency on the relative velocity $v$ between the annihilating particles; it is high as 10 pb when $v$ was important at the epoch of freeze out (in order to fulfill the dark-matter density estimate at this epoch by WMAP ($\Omega_{dm}h^2\simeq$0.1, \cite{spergel2003}), while being in the 10$^{-4}$ to 10$^{-5}$ pb range at present in the Galactic halo after the decrease of $v$ due to the expansion of the Universe.
	This LDMP annihilation scenario could involve a new neutral spin 1 boson $U$ (with a small coupling to ordinary matter), resulting from an additionally spontaneously broken $U(1)$ symmetry, for which a velocity dependent annihilation cross section can be accounted for naturally \citep{fayet2004}.


\section{Conclusions}

	Positron annihilation in the Galactic center region at a rate of the order of 1.3$\times$10$^{43}$ e$^+$ s$^{-1}$, compatible in morphology with the Galactic bulge, has recently been detected by INTEGRAL/SPI.
	Reviewing the astrophysical candidates capable of injecting the observed number of positrons into the Galactic bulge, we ruled out SNe~Ia as dominant injectors, due to their low explosion rate.
	Alternatively we have proposed SN2003dh-like hypernovae, capable of injecting 25 times more positrons than a typical SN~Ia.
	Hypernovae are expected to occur in the active star forming region in the Galactic center.
	Very first estimates (which will be refined with more observations) show however that their event rate might be too low as well, in order to be compatible with the steady-state $e^+$ production/annihilation hypothesis.
	In the lack of astrophysical solutions, light dark matter particles, concentrated in the Galactic central region and annihilating into $e^+ e^-$, have been proposed as a possible $e^+$ source.
	With more SPI data to come, the morphology of the 511 keV emission can be compared with the dark matter density profile, however still poorly known.
	Additionally, one has to bear in mind that the 511 keV-profile holds also the imprint of the gas distribution onto which the $e^+$ annihilate, which is poorly constrained as well.
	Furthermore, positrons - produced by light dark matter annihilation or hypernovae - could escape the production region and fill up the entire Galactic bulge, in which case the 511 keV-map would trace only the medium onto which the $e^+$ annihilate, and smear out completely the location of the $e^+$ sources.
	In order to confirm the dark matter hypothesis, searches for 511 keV emission from dark matter cusps in the halo of the Galaxy are ongoing, e.g. towards the Sagittarius Dwarf galaxy \citep{hooper2003,cordier2004}.

	The authors would like to express their gratitude to Pierre Fayet for very fruitful discussions on the subject of light dark matter.





\end{document}